\documentclass[prl,twocolumn,showpacs,preprintnumbers,amsmath,amssymb,floatfix]{revtex4}

\usepackage{subfigure,graphicx,epsfig,amsmath,amsfonts,amssymb,xcolor,slashed,ulem}

\newcommand{\be}{\begin{equation}}
\newcommand{\ee}{\end{equation}}
\newcommand{\ba}{\begin{eqnarray}}
\newcommand{\ea}{\end{eqnarray}}
\newcommand{\nn}{\nonumber}
\newcommand{\mev}{\textrm{ MeV}}
\newcommand{\JP}{J/\psi}
\newcommand{\decay}{\Lambda_b\to J/\psi\,K^- p}

\begin{document}

\title{ The LHCb pentaquark as a $\bar{D}^*\Sigma_c-\bar{D}^*\Sigma_c^*$ molecular state}
 
\author{L. Roca}
%\email{}
\affiliation{Departamento de F\'isica, Universidad de Murcia, E-30100 Murcia, Spain}

\author{J. Nieves}
%\email{}
\affiliation{IFIC, Centro Mixto Universidad de Valencia-CSIC,
Institutos de Investigaci\'on de Paterna,
Aptdo. 22085, 46071 Valencia, Spain}

\author{E.~Oset}

\affiliation{Departamento de
F\'{\i}sica Te\'orica and IFIC, Centro Mixto Universidad de
Valencia-CSIC Institutos de Investigaci\'on de Paterna, Aptdo.
22085, 46071 Valencia, Spain}

\date{\today}

\begin{abstract}

We perform a theoretical analysis of the $\Lambda_b \to J/\psi K^- p$ reaction from where a recent LHCb experiment extracts a $\Lambda(1405)$ contribution in the $K^- p$ spectrum close to threshold and two baryon states of hidden charm in the $J/\psi\,p$ spectrum. We recall that baryon states of this type have been theoretically predicted matching the mass, width and $J^P$ of the experiment, concretely some states built up from the $J/\psi\, N$, $\bar D^* \Lambda_c$, $\bar D^* \Sigma_c$, $\bar D \Sigma^*_c$ and $\bar D^* \Sigma^*_c$ coupled channels. We assume that the observed narrow state around 4450 MeV has this nature and we are able to describe simultaneously the shapes and relative strength of the the $K^- p$ mass distribution close to threshold and the peak of the $J/\psi\,p$ distribution, with values of the $J/\psi\, p$ coupling to the resonance in line with the theoretical ones. The non trivial matching of many properties gives support to a  $J^P=3/2^-$ assignment to this state and to its nature as a molecular state mostly made of   $\bar D^* \Sigma_c$ and $\bar D^* \Sigma^*_c$.

\end{abstract}

\maketitle

%\section{Introduction}  

The advent of chiral Lagrangians to account for the hadron-hadron interaction at low energies, and the further implementation of unitarity in coupled channels to go to higher energies, has allowed us to gain much insight into hadron interactions and shown that in many cases the interaction is strong enough to produce bound states in some channels, that decay into open ones of lesser strength, giving rise to known resonances or prediction of new ones  \cite{wolfram,npa,ramonet,angels,ollerulf,carmen,hyodo}. The introduction of the local hidden gauge Lagrangian \cite{hidden2,hidden4}, extending the chiral Lagrangians to account for the interaction of vector mesons, has allowed expansion of the field of research and provided new insight into mesonic \cite{raquel,geng} and baryonic states \cite{angelsvec,javi,kanchan}. In particular its extension to the heavy flavor sector has led to an adequate framework to investigate systems and predict states with open charm or beauty which match existing states \cite{opencharm,openbeauty}. Similarly, the same models have allowed us to make predictions for baryons with hidden charm or hidden beauty
\cite{wuhcharm,wuhbeauty,hiddencharm,hiddenbeauty,Wu:2010vk}, although in this case the lack of experimental information does not allow one to fine tune some parameters of the theory and one must live with larger uncertainties. In this sense, the predictions for the masses in these different approaches differ in about 100 MeV. The differences can be even bigger with the work of \cite{Garcia-Recio:2013gaa}, where the authors use an extended SU(8) spin-flavour symmetry Weinberg-Tomozawa interaction as the leading contribution, with modifications done to respect Heavy Quark Spin Symmetry (HQSS), and a particular renormalization/regularization scheme. Indeed, the bulk of the mentioned differences can be attributed precisely to the renormalization scheme, as discussed in \cite{hiddencharm}. Calculations with quark models and five quarks \cite{Yuan:2012wz} also lead to bound states of hidden charm, although they have a larger span of uncertainties. 

 Although baryons with open charm and open beauty have already been found, the recent experiment of \cite{exp} that finds a neat peak in the $J/\psi\,p$ invariant mass distribution from the $\Lambda_b \to J/\psi K^- p$ decay, is the first one to show signatures of a hidden charm baryon state. Although two states are reported from the $J/\psi\, p$ invariant mass distribution, the first one, at lower energies, is quite broad and one does not see any peak in that distribution. However, the hidden charm state around 4450~MeV,
called pentaquark $P_c(4450)^+$  in the experimental work \cite{exp},
shows up as a clear peak in this distribution, with a width of about 39$\pm5\pm19$ MeV, and this is the state we shall consider.  We shall take the work of \cite{hiddencharm} as reference. We find there, in the $I=1/2$ sector, one state  of $J^P=3/2^-$ mostly made of $\bar{D}^*\Sigma_c$ at 4417 MeV, with a width of about 8 MeV, which has a coupling to $J/\psi\, N$, $g=0.53$, and another one, mostly made of $\bar{D}^*\Sigma_c^*$ at 4481 MeV and with a width of about 35 MeV, which has a coupling to $J/\psi\, N$, $g=1.05$. The $3/2^-$ signature is one of the possible spin-parity assignments of the observed narrow state and its mass falls between these two predictions, although one must take into account that a mixture of states with $\bar{D}^* \Sigma_c$ and  $\bar{D}^*\Sigma^*_c$ is possible according to \cite{Garcia-Recio:2013gaa,uchinohc}. 
Although a preference for a $3/2^-$, $5/2^+$ assignment for the broad and narrow peaks, respectively, is indicated in the experimental work of \cite{exp}, the fact is that the $5/2^+$, $3/2^-$ identification is also statistically acceptable, since the likelihood of both assignments is very similar. Even other assignments are not discarded in the experimental paper.

 On the other hand, following similar steps to those in \cite{weihong} to interpret weak decays of heavy hadrons into light ones, a recent work has made predictions for the  $\Lambda_b \to J/\psi K^- p$ reaction and more concretely, $\Lambda_b \to J/\psi \Lambda(1405)$ \cite{rocamai}. Interestingly, the work of \cite{exp} also sees a bump in the $K^- p$ invariant mass distribution just above the  $K^- p$ threshold which is interpreted as due to $\Lambda(1405)$ production. 

  The purpose of the present work is to combine the information obtained from the experiment on the 
$K^- p$ invariant mass distribution close to threshold and the strength of the peak in the $J/\psi\, p$ spectrum and compare them to the theoretical results that one obtains combining the  results of \cite{hiddencharm} and \cite{rocamai}. As we shall see, we find a  $K^- p$ invariant mass distribution above the  $K^- p$ threshold, mainly due to the $\Lambda(1405)$, which agrees qualitatively with experiment, and the strength of this distribution together with the coupling that we find for the theoretical hidden charm state, produces a peak in the $J/\psi\,p$ spectrum, which agrees with the one reported in the experiment. These facts together provide support to the idea that the state found could be a hidden charm molecular state of  $\bar{D}^*\Sigma_c-\bar{D}^*\Sigma_c^*$ nature, predicted before by several theoretical groups.

%%%%%%%%%%%%%%%%%%%%%%%%%%%%%%%%%%%%%%%%%%%%%%%%%%%%%%%%%%%%%%
%%%%%%%%%%%%%%%%%%%%%%%%%%%%%%%%%%%%%%%%%%%%%%%%%%%%%%%%%%%%%%
%\section{Formalism}

\begin{figure}[tbp]
     \centering
     \subfigure[]{
          \label{fig:diag1a}
          \includegraphics[width=.7\linewidth]{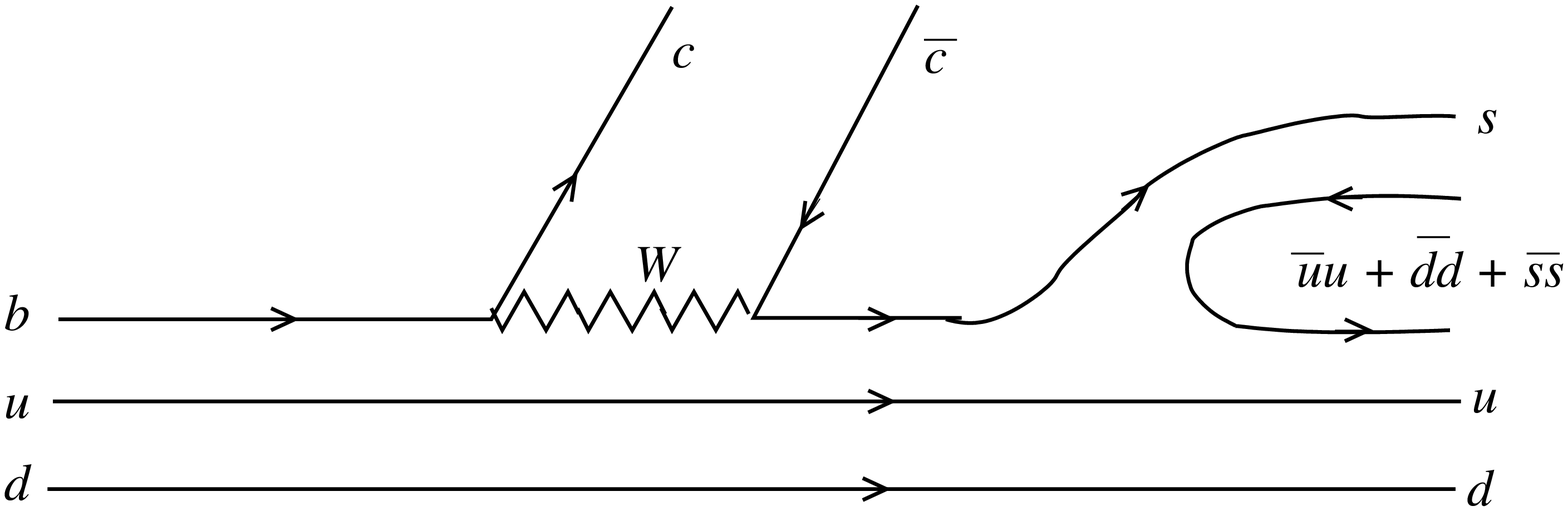}}\\
     \subfigure[]{
          \label{fig:diag1b}
          \includegraphics[width=.45\linewidth]{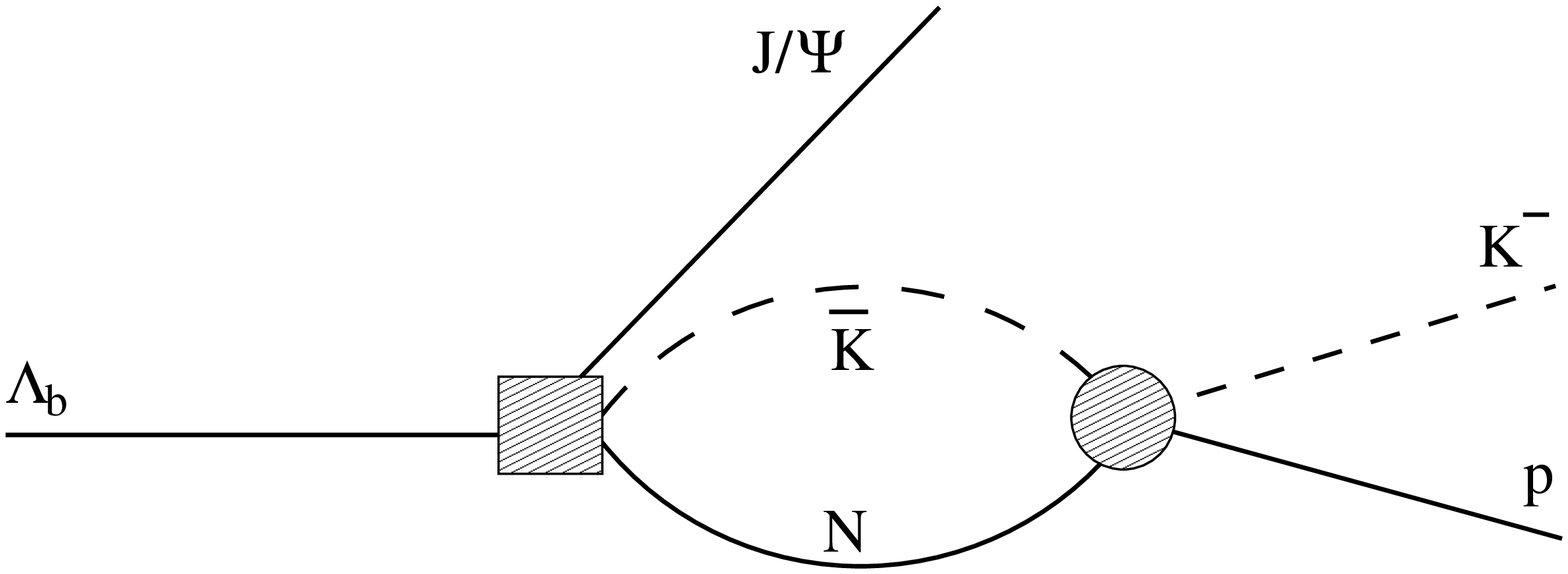}}
      \subfigure[]{
          \label{fig:diag1c}
          \includegraphics[width=.45\linewidth]{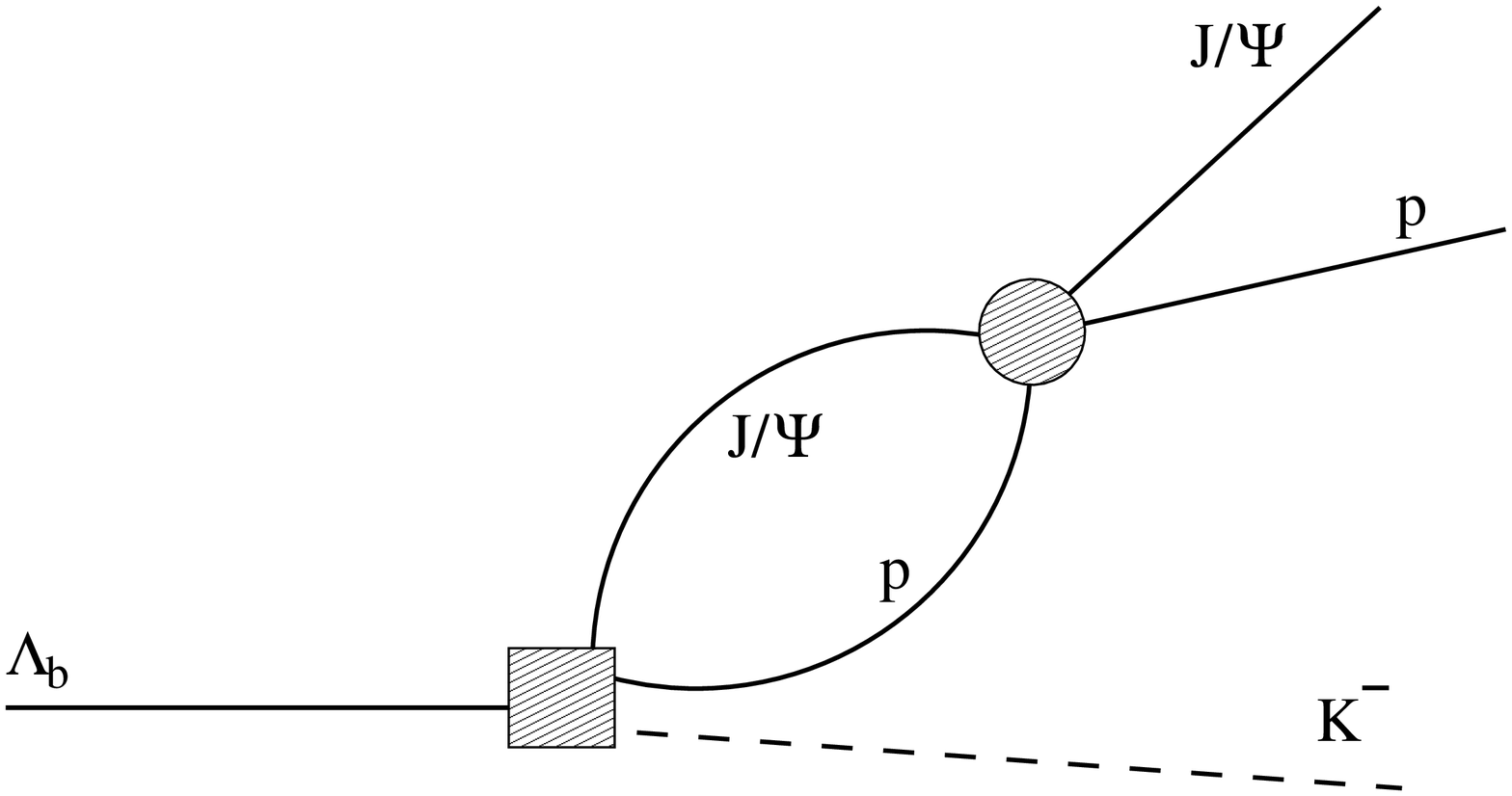}}
    \caption{Mechanisms for the $\Lambda_b\to J/\psi K^-p$ reaction implementing the final state interaction}\label{fig:diags}
\end{figure}

In ref.~\cite{rocamai} it was shown that the relevant mechanisms 
for the $\Lambda(1405)$ production in the 
$\decay$ decay 
are those depicted in 
Figs.~\ref{fig:diag1a} and \ref{fig:diag1b}.
Fig.~\ref{fig:diag1a} shows the basic process to produce a $K^-p$ pair from the weak decay 
of the 
${\Lambda_b}$. The $b$ quark in the $\Lambda_b$ first produces a $s$ quark and a
$c \bar c$ state,
which makes up the  
$J/\psi$, and then the three remaining quarks $u$, $d$ and $s$ hadronize into a meson-baryon pair. The $u$ and $d$ quarks remain as spectators in the process and carry isospin $I=0$, as in the initial state, producing, together with the $s$ quark, an isoscalar baryon after the weak process, and hence a meson-baryon system in $I=0$ after the hadronization of the $sud$ state. The  final meson-baryon state then undergoes final state interaction in coupled channels, Fig.~\ref{fig:diag1b}, from where the $\Lambda(1405)$ is 
dynamically produced.
Therefore the contribution to the $\Lambda_b\to J/\psi\,K^- p$ amplitude from the $\Lambda(1405)$ resonance, Figs.~\ref{fig:diag1a} and \ref{fig:diag1b}, is given by (see ref.~\cite{rocamai} for more details):

\ba\label{eqn:MKN}
T^{(K^-p)}(M_{K^-p})&=&V_p\bigg( h_{K^-p}+ \\ \nn
 &+& \sum_{i}h_iG_i(M_{K^-p})\,t_{i\,K^-p}(M_{K^-p}) \bigg)\,,
\ea
where $M_{K^-p}$ is the $K^-p$ invariant mass,
$h_i$ are numerical $SU(3)$ factors relating the production of the different meson-baryon channels $i$ in the hadronization (which explicit value and derivation can be found in ref.~\cite{rocamai}),
and $V_p$ accounts for CKM matrix elements and kinematic  prefactors.
In \cite{rocamai} it was found that the weights for primary production of the different meson-baryon channels after the weak process are
\begin{align*}
&h_{\pi^0\Sigma^0}=h_{\pi^+\Sigma^-}=h_{\pi^-\Sigma^+}=0\,,~h_{\eta\Lambda}=
-\frac{\sqrt{2}}{3}\,,\\
&h_{K^-p}=h_{\bar K^0n}=1\,,~h_{K^+\Xi^-}=h_{K^0\Xi^0}=0\,.
\end{align*}
The normalization of the amplitude in Eq.~\eqref{eqn:MKN} is such that the $K^-p$ invariant mass distribution is given by
\begin{align}\label{eqn:dGammadM}
\frac{d\Gamma_{K^-p}}{dM_{K^-p}}(M_{K^-p})
=\frac{1}{(2\pi)^3}\frac{M_p}{M_{\Lambda_b}}p_{J/\psi}\, p_K\left|T^{(K^-p)}\right|^2\,,
\end{align}
where $p_{J/\psi}$ and $p_K$ denote the modulus of the 
three-momentum
of the $J/\psi$ in the $\Lambda_b$ rest frame and that of the $K^-$ in the CM frame of the $K^-p$ pair.

Since we do not need the absolute normalization of the invariant mass distributions in the present work, the value of $V_p$ can be taken to be appropriate, as will be explained below when discussing the results. 
In Eq.~\eqref{eqn:MKN},
$G_i$ represents the meson-baryon loop function and 
$t_{ij}$ stands for the s-wave meson-baryon unitarized  scattering amplitudes from ref.~\cite{Roca:2013cca}.
 Note that the $\Lambda(1405)$ is not included as an explicit degree of freedom but it appears dynamically in the highly non-linear dynamics involved in the unitarization procedure leading to the $t_{ij}$ amplitudes. Actually two poles are obtained for the $\Lambda(1405)$ resonance
at $1352-48i$~MeV 
and $1419-29i$~MeV \cite{Roca:2013cca}. The highest mass $\Lambda(1405)$,
coupling mostly to $\bar K N$, is the one of relevance in the present work since it is closer to the energy region of concern.

It is interesting to note that Eq.~\eqref{eqn:MKN} explicitly includes the $J/\psi\,K^-p$ tree level contribution ($h_{K^-p}$ term in the parenthesis of Eq.~\eqref{eqn:MKN}), and that it interferes at the amplitude level with the second term in Eq.~\eqref{eqn:MKN} which, being proportional to the $t_{i\,K^-p}$ amplitudes, contains the effect of the $\Lambda(1405)$. However, the tree level contribution and its interference with the $\Lambda(1405)$ amplitude were not considered in the experimental analysis carried on in \cite{exp}. It is unclear what the effect of a reanalysis of the data including these contributions would be, but it might be worth trying.

On the other hand, in refs.~\cite{hiddencharm,Wu:2010vk},
 it was shown that the $J/\psi\,N$ final state interaction in coupled channels, considering also the $\bar D^* \Lambda_c$, $\bar D^* \Sigma_c$, $\bar D \Sigma^*_c$ and $\bar D^* \Sigma^*_c$, produces poles in the $J^P=3/2^-$, $I=1/2$, sector at
$4334+19i\mev$, $4417+4i\mev$ and $4481+17i\mev$, which couple sizeably to $\JP\,p$ (see table II in ref.~\cite{hiddencharm}).
 Therefore we can expect to see a resonance shape in the $J/\psi\, p$ invariant mass distribution in the $\Lambda_b\to J/\psi\,K^- p$ decay, maybe a mixture of the different poles. The mechanism for the final 
 $\JP\,N$ state interaction is depicted in Fig.~\ref{fig:diag1c}.
The filled circle in that figure represents the final $\JP\,p \to\JP\, p$ unitarized scattering amplitude. 
 Since in the real axis the shape of this amplitude  is very close to a Breit-Wigner form \cite{hiddencharm}, for the numerical evaluation in the present work we can effectively account for it by using

%In ref.~\cite{Xiao:2013yca} this amplitude is evaluated within the %chiral unitary approach implementing in the unitarization kernel %dynamics of the local hidden gauge Lagrangians extrapolated to $SU(4)$ %which were shown to be consistent to the constrains imposed
%by Heavy Quark Spin Symmetry.

\noindent
\be
t_{\JP\,p \to \JP\,p}= \frac{g^2_{\JP\,p}}{M_{\JP\,p}-M_R+i \frac{\Gamma_R}{2}}
\label{eq:tJpsi}
\ee
where $M_{\JP\,p}$ is the $\JP\,p$ invariant mass and $M_R$ ($\Gamma_R$) the mass (width) of the $P_c(4450)^+$. The amplitudes in refs.~
\cite{hiddencharm,Wu:2010vk} provide poles from where $M_R$ and $\Gamma_R$ can be directly obtained, but we fine tune these values to the  experimental results of ref.~\cite{exp}, $M_R=4449.8\mev$ and $\Gamma_R=40\mev$, which lie indeed in between the two heaviest poles obtained in 
ref.~\cite{hiddencharm}, as quoted above.
In Eq.~\eqref{eq:tJpsi}, $g_{\JP\,p}$ stands for the coupling
 of the dynamically generated resonance to $\JP\,p$, for which 
 a range of values from about 0.5 to 1 are obtained in refs.~\cite{hiddencharm,Wu:2010vk}, which are genuine 
 predictions of the theory.

The contribution of the $\JP\,p$ final state interaction to the amplitude is then

\ba
T^{(\JP\,p)}(M_{\JP\,p})&=&V_p\, h_{K^- p}G_{\JP\,p}(M_{\JP\,p}) \nn \\
&\times&t_{\JP\,p \to \JP\,p}(M_{\JP\,p})\,,
\label{eqn:MJpsip}
\ea
with $G_{\JP\,p}$ the $\JP\,p$ loop function regularized by dimensional regularization as in ref.~\cite{hiddencharm}.

Since the main building blocks of the $P_c(4450)^+$ state in \cite{hiddencharm} are 
 $\bar D^* \Sigma_c$ and  $\bar D^* \Sigma^*_c$, in principle the main sequence to produce this baryon should be
of the type $\Lambda_b\to K^- \bar D^* \Sigma^*_c \to K^- p J/\psi$ (the argument that follows hold equally for $\Sigma_c$), where one produces  $K^- \bar D^* \Sigma^*_c$ in the first step and the  $\bar D^* \Sigma^*_c \to  p J/\psi$ transition would provide the resonant amplitude accounting for the $P_c(4450)^+$ state in the $\JP\,p$ spectrum. The topology to produce the first step would be now different to that depicted in Fig.~\ref{fig:diag1a},
 corresponding to external emission in the classification of ref. \cite{chau}, with the quark sequence  $b \to W c$, the $W$ producing
 an initial $s\bar c$ pair, which upon hadronization will give rise to $\bar K \bar D^*$.
  On the other hand, as mentioned above, the $u d$ quarks are in $I=0$ in the initial state and act as spectators, hence conserve the $I=0$ character after the weak process and, thus, cannot lead to $\Sigma_c^*$ states.  One could form $\Lambda_c$ instead of $\Sigma_c^*$, and  $\bar D^* \Lambda_c$ is one of the channels coupling to $P_c(4450)^+$, but the coupling of this channel in \cite{hiddencharm} is very small (about five times smaller than that of $J/\psi$).
  There are other possibilities of hadronization, where one of the $u$, $d$ quarks present in the  $\Lambda_b$ do not act as spectators but instead recombine with the $\bar c$ antiquark to form a $\bar D^*$ meson. The other quarks can now, after hadronization, produce
 $\bar K\Sigma^*_c$ \cite{Wang:2015jsa}. However, this mechanism is suppressed, since it involves a large mismatch of momenta between that carried by the light quarks in the $\Lambda_b$ and the momentum of the outgoing $\bar D^*$.
 A more detailed discussion of this issue is given in 
 \cite{Miyahara:2015cja}.
     The $P_c(4450)^+$ resonance therefore has to be produced by rescattering of $J/\psi\,p$ after the primary production of $\Lambda_b \to J/\psi K^- p$ through the mechanism depicted in Fig. \ref{fig:diag1c}. This feature of the reaction is what allows us to relate the $P_c(4450)^+$ production with the $K^- p$ production, i.e, the factor $V_p\, h_{K^- p}$ enters the production of both the $\Lambda(1405)$, via Eq. (\ref{eqn:MKN}), and of the  $P_c(4450)^+$, via Eq. (\ref{eqn:MJpsip}).

%\section{Results}

\begin{figure}[tbp]
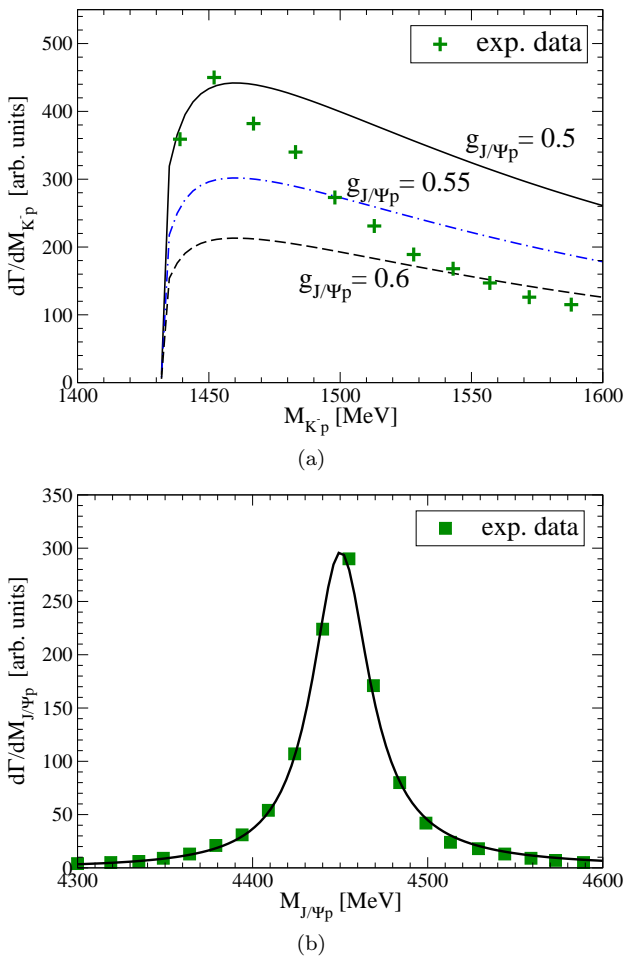

     \centering
     \subfigure[]{
          \label{fig:res1}
          \includegraphics[width=.95\linewidth]{figure2a.eps}}
     \subfigure[]{
          \label{fig:res2}
          \includegraphics[width=.95\linewidth]{figure2b.eps}}
    \caption{Results for the $K^- p$ and $\JP\,p$ invariant mass distributions compared to the data of ref. \cite{exp}.}
    \label{fig:res}
\end{figure}

In Fig.~\ref{fig:res} we show the results for the $K^- p$  and $\JP\,p$  invariant mass distributions compared to the experimental data of ref.~\cite{exp}. The absolute normalization is arbitrary but the same for both panels. In the data shown for the $K^- p$ mass 
distribution  only the $\Lambda(1405)$ contribution is included, i.e., it shows the result of the $\Lambda(1405)$ component of the experimental analysis carried out in \cite{exp}. Therefore, in order to compare to this data set, only the 
amplitude of Eq.~\eqref{eqn:MKN} is considered. Similarly, the experimental $\JP\,p$ mass distribution shown in 
Fig.~\ref{fig:res}(b) only considers the contribution from the $P_c(4450)^+$ and, thus, the theoretical calculation for 
Fig.~\ref{fig:res}(b)  only considers the amplitude of Eq.~(\ref{eqn:MJpsip}).
The different curves are evaluated considering different values for the coupling of the $P_c(4450)^+$  to $J/\psi\,p$, ($g_{\JP\,p}=0.5$, 0.55 and 0.6). For each value of $g_{\JP\,p}$, 
$V_P$ has been normalized such that the peak of the $\JP\,p$ distribution agrees with experiment, and this is why there is only one curve for the $\JP\,p$ mass distribution.
Note that the mass of the higher  $\Lambda(1405)$ resonance lies below the $K^- p$ threshold. Therefore the accumulation of strength close to threshold is due to the tail of that resonance. The experimental analysis of Ref.~\cite{exp} parametrizes the contribution of the $\Lambda(1405)$ as a Breit Wigner, which falls down fast above threshold as a function of the $K^- p$ invariant mass. In the chiral unitary approach used here, the resonance is dynamically generated and, in addition, a background (tree level term in Eq.~\eqref{eqn:MKN}) is also present. Altogether provides a milder slow down with the $K^- p$ invariant mass, that renders our distribution flatter than that obtained in the experimental analysis.

The results are very sensitive to the value of the $\JP\,p$ coupling since the $J/\psi\,p$ partial decay width is proportional to $g_{\JP\,p}^4$. The figure shows that a value for the coupling of about 0.5 can account fairly for the relative strength between the $ \JP\,p$ and $K^- p$ mass distributions. This value of the coupling is of the order
 obtained in the extended local hidden gauge unitary approach of refs.~\cite{hiddencharm,Wu:2010vk} which is a non-trivial consequence of the theoretical model since the value of this coupling is a reflection of the highly non-linear dynamics involved in the unitarization of the scattering amplitudes.

   It is also worth noting that the values of $g_{\JP\,p}$ used, lead to a partial decay width of $P_c(4450)^+$ into $\JP\,p$ ($\Gamma=M_N g_{\JP\,p}^2~p_{J/\psi}/(2\pi M_R)$) of 6.9 MeV, 8.3 MeV, 9.9 MeV, which are of the order of the experimental width, but smaller as it should be, indicating that this channel is one of the relevant ones in the decay of the $P_c(4450)^+$ state.

Reproduction of the exact shape of the experiment in Fig.~\ref{fig:res}(a) is not essential here. First, there are assumptions made in the experiment to extract the $\Lambda(1405)$ signal as mentioned above. Furthermore, one should take into account that different theoretical models that equally well fit the $K^- p \to X$ data, lead to  different shapes for this $K^- p$ distribution. In fact, as seen in \cite{rocamai}, the Bonn model of \cite{maimeissner} produces a shape with a faster fall down, more in agreement with the experiment.  
We emphasize the fact that the chiral unitary approach provides a full amplitude and not only the resonance contribution. Hence, the comparison with ref. \cite{exp} is meaningful for values of $M_{inv}$ close to threshold, about 50 MeV above threshold.

The fact that we can fairly reproduce the relative strength of the mass distributions with values of the coupling in the range predicted by the coupled channels unitary approach, provides  support to the  interpretation of the $P_c(4450)^+$ state as dynamically generated from the coupled channels considered and to the $3/2^-$ signature of the state.

We should nevertheless acknowledge that we do not have an explanation for the broad state $P_c(4380)$ claimed in the experimental analysis. The experimental support for the existence of this state is not as strong as for the narrow one. Indeed, its Argand plot is not as clean
 as that of the narrow $P_c(4450)$ \cite{exp}. On the other hand it is worth recalling that in spite of its large strength in the experimental analysis, it does not show up, even as a bump in the $\JP p$ experimental invariant mass distribution. One should also mention that this peak in the $\JP p$ mass distribution mostly accounts for the strength of the  $K^- p$ distribution at large invariant masses where background terms, in particular the tree level contribution advocated in the present work, might play a role. 

 The findings of \cite{exp} prompted the work of \cite{slzhu} where, using a boson exchange model \cite{slzmodel}, molecular structures of  $\bar D^* \Sigma_c$ and $\bar D^* \Sigma^*_c$ are also obtained with similarities to our earlier work of \cite{hiddencharm}. However, the interrelation between the $J/\psi\,p$ and $K^- p$ invariant mass distributions is not addressed in \cite{slzhu}.

It should also be mentioned that the possibility that the peak observed experimentally is just a kinematical effect due to a triangular singularity involving the
$\chi_{c1} p\to \JP p$
transition has been pointed out recently
\cite{Guo:2015umn}.
Since this transition amplitude is OZI forbidden it would be most appropriate to make reliable estimates of its strength in order to determine the relevance of this mechanism compared to the experimental findings.

Summarizing, it is a confluence of many facts that leads us to make these claims about the new state found: The fact that theoretically it was predicted with $3/2^-$ and the range of the mass and the value of the width agree with experiment. The fact that the $J/\psi\,p$ is only a minor channel in the build up of the resonance (the dominant channels have a coupling of the order of 3, compared to 0.5 of $J/\psi\,p$), but one relevant in the decay. The fact that 
the production of 
the dominant $\bar D^* \Sigma_c$ and $\bar D^* \Sigma^*_c$ components of $P_c(4450)^+$ is suppressed in the weak process, which forces the resonance to be produced from $J/\psi\,p$ rescattering. This was essential for us to be able to relate the $K^- p $ production with 
$J/\psi\,p$ production. Also the couplings $g_{\JP\,p}$ needed to match the two mass spectra are in line with the theoretical predictions and lead to a partial decay width into the $p J/\psi$ channel of the order, but smaller, than the total width of the $P_c(4450)^+$ state. 
The matching of these pieces in this puzzle supports the interpretation of the $P_c(4450)^+$ state found as a molecular state of mostly  $\bar D^* \Sigma_c$ and $\bar D^* \Sigma^*_c$ nature with isospin $I=1/2$ and spin parity $3/2^-$.

\section*{Acknowledgments}  

This work is partly supported
by the Spanish Ministerio de Economia y Competitividad and European
FEDER funds under contracts number FIS2011-28853-C02-01,
FIS2011-28853-C02-02, FIS2014-51948-C2-1-P, FIS2014-51948-C2-2-P, FPA2013-40483-P, FIS2014-57026-REDT and the Generalitat Valenciana in the program
Prometeo II-2014/068. We acknowledge the support of the European
Community-Research Infrastructure Integrating Activity Study of
Strongly Interacting Matter (acronym HadronPhysics3, Grant Agreement
n. 283286) under the Seventh Framework Programme of EU. 
This work is partially funded by the grants MINECO (Spain) and ERDF (EU), grant FPA2013-40483-P.

\end{document}